\documentclass{article}

\usepackage{microtype}
\usepackage{graphicx}
\usepackage{subcaption}
\usepackage{booktabs} 
\usepackage{multirow}  
\usepackage{makecell}  

\usepackage{hyperref}
\usepackage{url}

\usepackage[preprint]{icml2026}

\usepackage{amsmath}
\usepackage{amssymb}
\usepackage{mathtools}
\usepackage{amsthm}
\usepackage{enumitem}
\usepackage{balance}

\usepackage[capitalize,noabbrev]{cleveref}

\theoremstyle{plain}

\theoremstyle{definition}

\theoremstyle{remark}

\usepackage[textsize=tiny]{todonotes}

\icmltitlerunning{FastCode: Fast and Cost-Efficient Code Understanding and Reasoning}

\begin{document}

\twocolumn[
  \icmltitle{FastCode: Fast and Cost-Efficient Code Understanding and Reasoning}



  \icmlsetsymbol{equal}{*}

  \begin{icmlauthorlist}
    \icmlauthor{Zhonghang Li}{equal,HKU}
    \icmlauthor{Zongwei Li}{equal,HKU}
    \icmlauthor{Yuxuan Chen}{HKU}
    \icmlauthor{Han Shi}{HUAWEI}
    \icmlauthor{Jiawei Li}{HUAWEI}
    \icmlauthor{Jierun Chen}{HUAWEI}
    \icmlauthor{Haoli Bai}{HUAWEI}
    \icmlauthor{Chao Huang}{HKU}
  \end{icmlauthorlist}


  \icmlaffiliation{HKU}{The University of Hong Kong}
  \icmlaffiliation{HUAWEI}{Huawei Noah’s Ark Lab}

  \icmlcorrespondingauthor{Chao Huang}{chaohuang75@gmail.com}


  \vskip 0.3in
]

\def\model{FastCode}



\printAffiliationsAndNotice{}  

\begin{abstract}

Repository-scale code reasoning is a cornerstone of modern AI-assisted software engineering, enabling Large Language Models (LLMs) to handle complex workflows from program comprehension to complex debugging. However, balancing accuracy with context cost remains a significant bottleneck, as existing agentic approaches often waste computational resources through inefficient, iterative full-text exploration. To address this, we introduce \model, a framework that decouples repository exploration from content consumption. \model\ utilizes a structural scouting mechanism to navigate a lightweight semantic-structural map of the codebase, allowing the system to trace dependencies and pinpoint relevant targets without the overhead of full-text ingestion. By leveraging structure-aware navigation tools regulated by a cost-aware policy, the framework constructs high-value contexts in a single, optimized step. Extensive evaluations on the SWE-QA, LongCodeQA, LOC-BENCH, and GitTaskBench benchmarks demonstrate that \model\ consistently outperforms state-of-the-art baselines in reasoning accuracy while significantly reducing token consumption, validating the efficiency of scouting-first strategies for large-scale code reasoning. 
Source code is available at \url{https://github.com/HKUDS/FastCode}.
\end{abstract}

\section{Introduction}

LLMs are increasingly embedded in real-world software engineering workflows, from answering developer questions to assisting with debugging, refactoring, and feature development~\cite{wang2025swedev,li2025deepcode}. Many of these workflows fundamentally require \emph{repository-scale code reasoning}: the ability to interpret a natural language request, identify the relevant parts of a large codebase, trace cross-file dependencies, and synthesize a grounded response or actionable plan~\cite{peng2025sweqa,deshpande2024classlevel}. 
In modern engineering organizations, stronger repository-level understanding improves developer productivity and reduces costs by minimizing time navigating unfamiliar code and preventing avoidable regressions.

At the same time, repository-scale code reasoning is substantially more challenging than snippet-level understanding. It must reconcile local implementation details with the global architecture, which often involves multi-hop dependency tracing across files, symbols, and abstraction layers~\cite{chen2025locagent,liu2024graphcoder,liu2025codexgraph}. This difficulty surfaces in common and economically important tasks such as repository-level question answering (QA), code localization, and multi-hop comprehension for repair. In practice, \emph{token and compute budgets} limit how much of the repository can be inspected per query, making it difficult to gather enough relevant evidence for grounded answers without pulling in excessive and potentially irrelevant context.

Despite rapid progress, existing approaches still face a core bottleneck: \textbf{the tension between accuracy and context cost}. General LLMs struggle to reliably answer repository-level questions when they cannot see enough of the codebase. A naive solution is to feed more code into the prompt, but repository contexts quickly become prohibitively large, incurring substantial token cost, and even long-context models can be sensitive to irrelevant or noisy context. Standard retrieval-augmented generation (RAG) pipelines partially mitigate this by chunking and retrieving code as flat text, but this often severs structural relationships that are essential for reasoning---such as imports, inheritance hierarchies, and call chains. 
To improve grounding, recent approaches such as DeepWiki~\cite{deepwiki2025} and CodeWiki~\cite{hurley2025codewiki} construct persistent repository-wide indices or synthesize documentation artifacts offline. Although effective for long-lived deployments, this strategy front-loads significant overhead to the indexing stage, incurring substantial time and token costs that are disproportionate to justify for short-lived or ad-hoc usage. 

Agentic approaches aim to overcome these limitations by iteratively exploring the repository (e.g., searching, opening files, and refining hypotheses). Recent code agents adopt varying interface designs to facilitate such exploration: SWE-agent~\cite{yang2024sweagent} employs custom agent-computer interfaces with specialized file navigation and editing commands, while systems like OpenHands~\cite{wang2025openhands}, Cursor~\cite{cursor2026} and Claude Code~\cite{anthropic2026claudecode} leverage general-purpose code execution (bash, Python) as their primary interaction mechanism. 
Despite these architectural differences, all approaches fundamentally rely on sequential, token-intensive repository traversal to locate relevant code contexts. Yet in repository-scale settings, such exploration can become expensive and brittle: each additional ``read'' consumes substantial tokens, and iterative loops amplify cost as the agent re-ingests the accumulating context history at each step. In practice, the system often pays a high token price to discover that many inspected files are irrelevant, or it terminates early and produces answers that are insufficiently grounded.

\begin{figure}[t]
    \centering
    \begin{subfigure}[b]{0.23\textwidth}
        \centering
        \includegraphics[width=\textwidth]{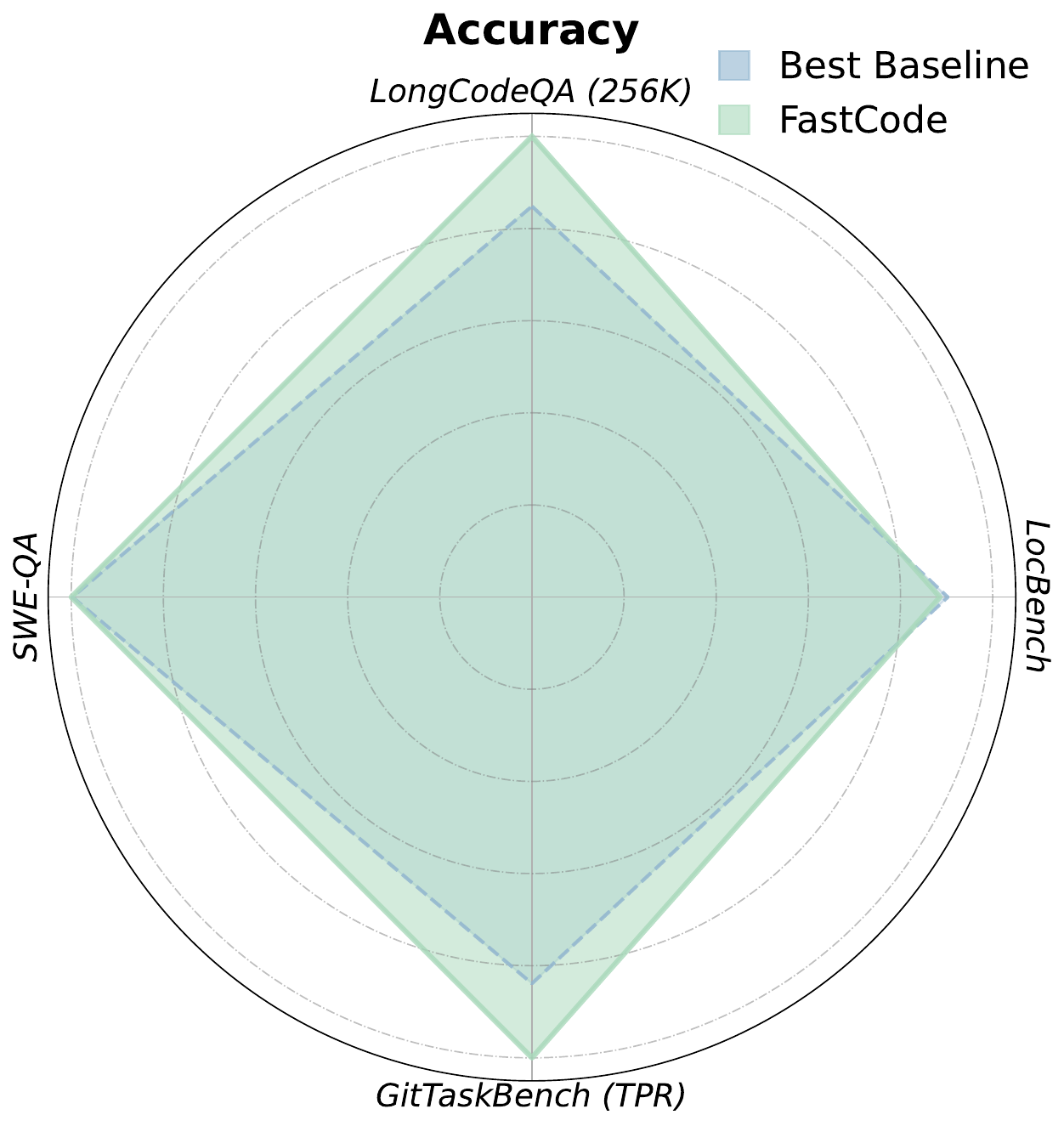}
    \end{subfigure}
    \hfill
    \begin{subfigure}[b]{0.23\textwidth}
        \centering
        \includegraphics[width=\textwidth]{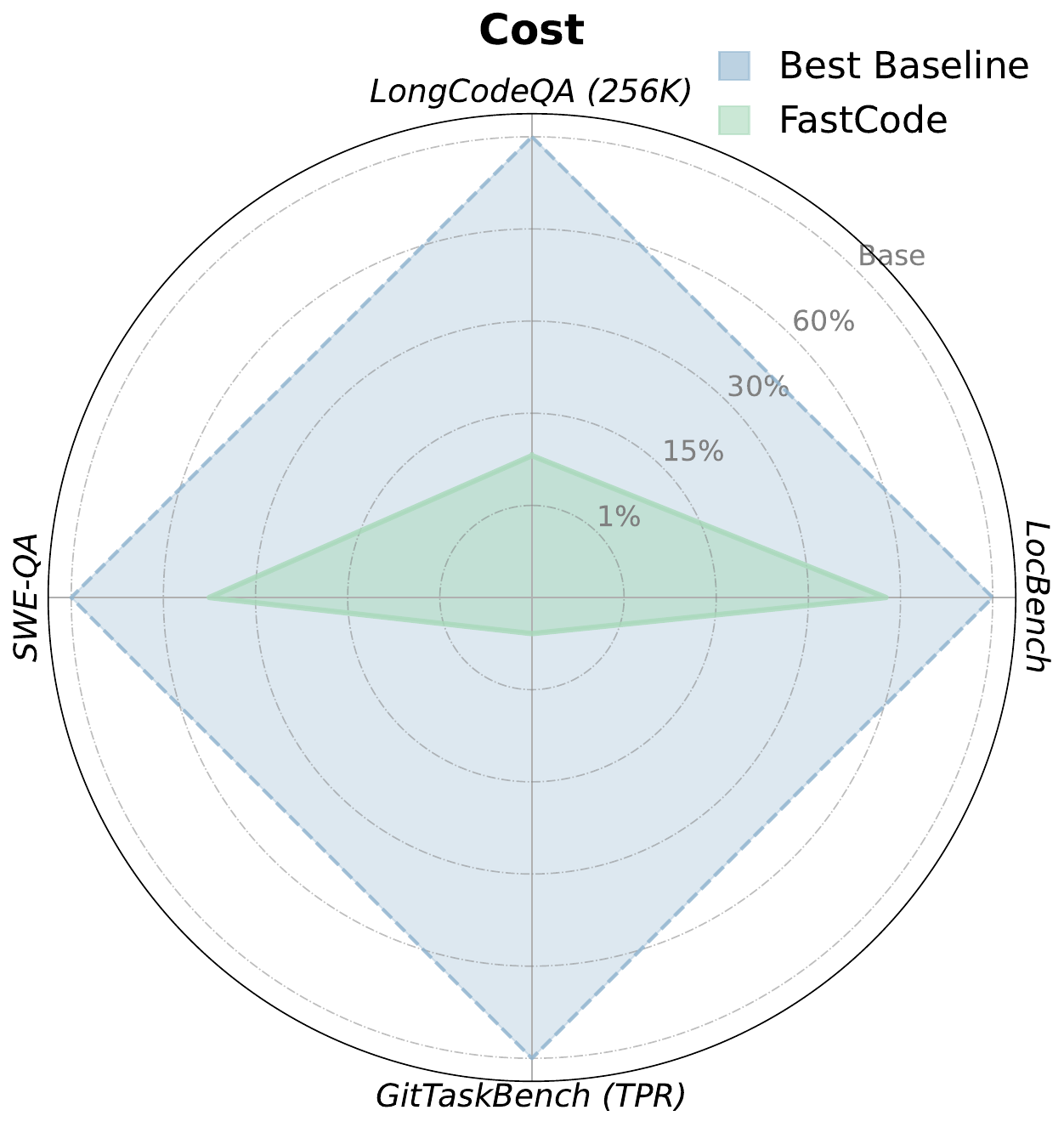}
    \end{subfigure}
    \vspace{-0.05in}
    \caption{Accuracy and cost trade-off comparison}
    \label{fig:tradeoff}
    \vspace{-0.18in}
\end{figure}

To balance reasoning accuracy with computational cost, we introduce \model, a framework that decouples repository exploration from content consumption. To circumvent the prohibitive token overhead of iterative file reading, \model\ employs a structural scouting mechanism: it navigates the codebase using lightweight metadata to pinpoint relevant targets before loading content, committing to full-text ingestion only to construct a high-value context in a single, cost-effective step. Our approach orchestrates this process through three synergistic components: 
(1) a semantic-structural representation method that provides the \emph{scouting map}, unifying lightweight metadata with a multi-layered dependency graph to enable visibility without reading; 
(2) a codebase context navigation mechanism that performs the \emph{exploration}, utilizing structure-aware tools to actively prune the search space and identify high-potential candidates; and 
(3) a cost-aware context management policy that makes the \emph{selection decision}, dynamically optimizing the final context assembly to maximize reasoning confidence using the minimum sufficient context.

We evaluate \model\ on challenging repository-level benchmarks, including SWE-QA~\cite{peng2025sweqa}, LongCodeQA~\cite{rando2025longcodebench}, LocBench~\cite{chen2025locagent}, and GitTaskBench~\cite{ni2025gittaskbench}. Experimental results demonstrate that \model\ consistently outperforms state-of-the-art baselines in reasoning accuracy while significantly reducing token consumption. By effectively pruning irrelevant context through structural scouting, our framework validates that high-precision repository reasoning is achievable without the prohibitive computational costs associated with standard iterative retrieval. The main contributions are:

\begin{itemize}[leftmargin=*]
    \vspace{-0.1in}
    \item We introduce a \emph{scouting-first} code reasoning paradigm that leverages a semantic-structural map to decouple exploration from consumption. By unifying lightweight metadata with dependency graphs, this approach enables the system to pinpoint and navigate to the most relevant code elements prior to retrieving full file content.
    \vspace{-0.05in}
    \item We develop a structure-guided navigation strategy regulated by a cost-aware control policy. This framework localizes targets by actively tracing topological dependencies and intent cues, while simultaneously optimizing context cost through a dynamic mechanism that evaluates the incremental value of context expansion to prune irrelevant paths and terminate efficiently.
    \vspace{-0.05in}
    \item Extensive experiments on four repository-level benchmarks validate the practical efficacy of the proposed framework. The results confirm that our framework improves repository-level reasoning accuracy and achieves substantial token savings, demonstrating the value of structural scouting for resource-constrained applications.
\end{itemize}
\section{Preliminary}

\subsection{Repository Representation}
We formalize the target repository $\mathcal{R}$ as a \textbf{semantic map}, structured as a directed graph $\mathcal{G} = (\mathcal{U}, \mathcal{E})$. Here, $\mathcal{U}$ represents the discrete landscape of hierarchical code units (e.g., files, classes), while $\mathcal{E} = \{G_{\mathrm{dep}}, G_{\mathrm{inh}}, G_{\mathrm{call}}\}$ denotes the topological pathways capturing dependencies, inheritance, and call flows. Crucially, each unit $u \in \mathcal{U}$ serves as a navigable landmark, comprising \textit{lightweight metadata} (e.g., signatures) for rapid scouting and a \textit{full implementation body} for deep reasoning. This dual representation enables the agent to explore the map and assess relevance without the computational overhead of traversing full code bodies.

\subsection{Task Objective}
Given a natural language query $q$ and the structured repository $\mathcal{G}$, our goal is to generate a solution $y$ (e.g., an answer or an executable patch) by navigating this map. Unlike static retrieval, we formulate the reasoning process as a joint optimization of relevance and efficiency. The system aims to identify an optimal context $\mathcal{C}^*$ that maximizes semantic sufficiency while penalizing token consumption:
\begin{equation}
    y = \mathcal{M}(q, \mathcal{C}^*), \; \mathcal{C}^* = \operatorname*{argmax}_{\mathcal{C} \subset \mathcal{U}} \big[ \operatorname{Rel}(\mathcal{C} \mid q) - \lambda \Omega(\mathcal{C}) \big]
\end{equation}
Here, $\mathcal{M}$ denotes the reasoning model that generates $y$ by processing the full content of $\mathcal{C}^*$. The optimization objective balances two competing goals: $\operatorname{Rel}(\cdot)$ quantifies the relevance of the selected context $\mathcal{C}$, while $\Omega(\cdot)$ is the computational overhead, modulated by a penalty coefficient $\lambda$.
\section{Methodology}
\begin{figure*}[t]
    \centering
    \includegraphics[width=\textwidth]{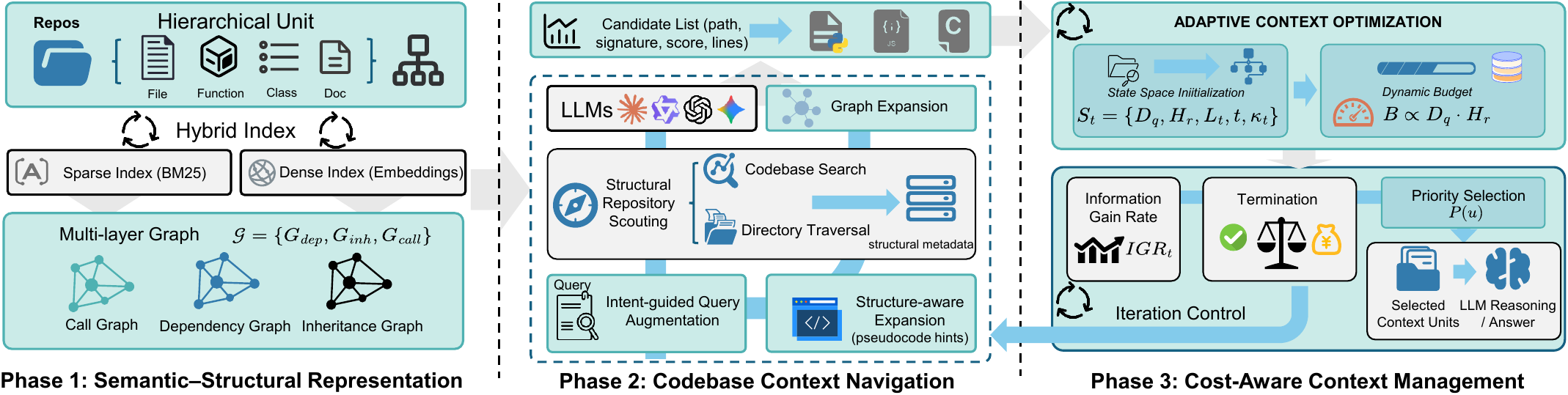}
    \caption{The overall framework of \model.}
    \label{fig:framework}
    \vspace{-0.1in}
\end{figure*}

To optimize the trade-off between reasoning accuracy and computational cost in repository-scale tasks, we propose \model, a cost-aware agentic framework that reformulates context acquisition as a structure-guided navigation process. As illustrated in Figure \ref{fig:framework}, our approach consists of three components: (1) a \textit{Code Semantic-Structural Representation} that unifies hybrid indexing with a multi-layered dependency graph; (2) a \textit{Codebase Context Navigation} mechanism that employs a scouting mechanism to explore repository layouts without incurring the cost of full-content reading; and (3) a \textit{Cost-Aware Context Management} policy that dynamically regulates information gathering to maximize reasoning confidence within a limited token budget.

\subsection{Semantic–Structural Code Representation}
\label{sec:representation}

Standard RAG approaches treat code as flat textual chunks, severing intrinsic structural dependencies like inheritance and imports. To bridge this semantic gap, \model\ constructs a holistic representation that preserves the codebase's topological integrity. By decomposing the repository into interconnected semantic units rather than isolated segments, our framework enables precise navigation from global architecture to local logic through multi-grained hybrid indexing and symbol-aware relation modeling.

\subsubsection{Hierarchical Code Units}
Optimal retrieval granularity varies with query intent: architectural reasoning demands file-level context, whereas debugging requires function-level specifics. To support this diversity, we parse the repository into a hierarchical tree spanning four levels: \textbf{File}, \textbf{Class}, \textbf{Function}, and \textbf{Documentation}. Crucially, instead of indexing only raw implementations, we extract lightweight metadata for each node—such as type signatures, docstrings and line ranges. This constructs a token-efficient skeletal view of the repository, enabling the subsequent agent to evaluate relevance without incurring the computational cost of reading full code bodies.

\subsubsection{Multi-Grained Hybrid Indexing}
Code retrieval demands dual capabilities: pinpointing exact identifiers and understanding abstract intent. To reconcile these requirements, we implement a Dual-Index Mechanism that synergizes sparse and dense representations:

\textbf{(i) Sparse Index (Lexical):} Leveraging BM25, we target precise code tokens (e.g., function signatures) to preserve explicit keyword matches often diluted in vector spaces.

\textbf{(ii) Dense Index (Semantic):} We employ a embedding model to map multi-grained units into high-dimensional vectors. This captures implicit semantics, identifying synonymous implementations even without lexical overlap.

\subsubsection{Symbol-Aware Relation Modeling}

To mitigate context fragmentation across file boundaries, we construct a code relationship graph using efficient Abstract Syntax Tree (AST)-based analysis, avoiding the overhead of heavy compiler infrastructure. We formalize the repository as a multi-layered directed graph $\mathcal{G} = \{G_{dep}, G_{inh}, G_{call}\}$, capturing three distinct topological dimensions:

\begin{itemize}[leftmargin=*]
    \vspace{-0.10in}
    \item \textbf{Dependency Layer ($G_{dep}$):} Maps file-level import relationships, where edges represent module dependencies.
    \item \textbf{Inheritance Layer ($G_{inh}$):} Maps class hierarchies, where directed edges denote base-class derivation.
    \item \textbf{Call Layer ($G_{call}$):} Maps execution flow, linking function invocation sites to their precise definitions.
\end{itemize}


\subsection{Codebase Context Navigation}
\label{sec:navigation}
Precise localization of query-related code elements is a prerequisite for effective reasoning, yet it is hindered by a dual challenge: the semantic ambiguity of natural language queries and the prohibitive token cost of the iterative retrieval-and-reading loop. Deciding whether a source file is relevant often requires reading its content, which rapidly exhausts context windows. To address this, we introduce a navigation phase that bridges the semantic gap and performs low-cost scouting, efficiently pruning the search space before committing to expensive content reading.

\subsubsection{Query Augmentation}
\label{sec:query_aug}
To guide navigation effectively, the system first categorizes the query's intent to orchestrate a dual-track enhancement strategy. First, the query is rewritten to align with the identified intent, optimizing the semantic retrieval for abstract concept matching. Simultaneously, we perform targeted keyword expansion, extracting domain-specific terminology and API candidates to boost the BM25 retrieval.
This separation ensures that semantic vectors capture high-level logic while lexical indices identify precise symbol definitions.

While keyword expansion captures terminology, it misses algorithmic structure. We employ the LLM to generate pseudocode hints--skeletal implementations with specific library calls, control flows, or class hierarchies. These hints serve as structural anchors, enabling retrieval based on logic similarity and API usage rather than textual overlap, enhancing recall by aligning the query's latent logic with the codebase's execution flow.

\subsubsection{Structural Repository Scouting}
\label{sec:scouting}

To address the overhead of iteratively reading code bodies in existing coding agent, we propose a structural scouting strategy: the agent assesses candidate value through lightweight metadata, decoupling exploration from consumption and allocating expensive context window resources only to high-valued targets.

\vspace{-0.1in}
\paragraph{Tool-Assisted Exploration}
We provide the agent with the following two code navigation tools:
\begin{itemize}[leftmargin=*]
    \vspace{-0.1in}
    \item \emph{Directory Traversal:}  Lists files under a specified directory to identify relevant modules.
    \item \emph{Codebase Search:} A regex-enhanced scanner that returns file path and match counts and without loading full files.
    \vspace{-0.1in}
\end{itemize}
Tool call results are parsed into structural metadata (e.g., file paths, class/function signatures) rather than full implementation details.  The agent uses these lightweight signals to filter candidates—prioritizing files with high match density and relevant signatures while discarding irrelevant ones.

\vspace{-0.1in}
\paragraph{Adaptive Scouting Workflow}
We structure the navigation as an iterative decision process over rounds $t \in \{1, \dots, T\}$.

\textbf{Initialization (Round 1):} The system assesses the user query and adaptively generates up to two distinct streams:
\begin{itemize}[leftmargin=*]
    \vspace{-0.1in}
    \item \emph{Alignment Parameters:} Refined intent and keywords to guide the Semantic/Lexical Retriever (in Section \ref{sec:query_aug}).
    \vspace{-0.05in}
    \item \emph{Exploratory Tool Calls:} Optional execution commands for the exploration tools if the query requires specific pattern matching (e.g., finding usage of a specific API).
    \vspace{-0.1in}
\end{itemize}

These two streams are complementary: retrieval identifies similar files, while tool-based exploration handles precise pattern matching and structural navigation.

\textbf{Graph Expansion:}
After deduplicating candidates from both streams, we apply graph extension to capture non-lexical dependencies: the system queries Call, Dependency, and Inheritance graphs to include logically related units, revealing structural relationships beyond initial matches.

\textbf{Iterative Refinement (Round $t$):}
The merged candidates are presented to the agent as a consolidated metadata list. For each candidate unit (File, Class, or Function level), the system exposes a multi-dimensional profile, including the: \textbf{(i) Provenance:} Source of discovery (retrieval, tools call, or graph neighbor) with corresponding relevance scores.
\textbf{(ii) Structural Identity:} Signatures, inheritance bases, docstrings, and relationships to existing candidates (for graph-expanded units).
\textbf{(iii) Cost Metric:} Total line count.
Based on this aggregated view, the agent decides which units to keep and whether to terminate scouting or issue new tool calls to resolve remaining uncertainty.

\subsection{Cost-Aware Context Management}
\label{sec:reasoning}

To balance accuracy and cost, \model\ treats context construction as a state-aware adaptive process. By monitoring the evolving reasoning state against the remaining budget, the framework dynamically regulates information gathering—deciding when to acquire details or terminate the search to ensure optimal context within resource limits.

\subsubsection{State Space Initialization}
\label{sec:state_init}

We formalize the iterative retrieval as a trajectory through a multi-dimensional state space. At each iteration step $t$, the agent observes a state vector $S_t = \{D_q, H_r, L_t, t, \kappa_t\}$, which synthesizes static environmental factors with dynamic runtime metrics to guide the policy $\pi(S_t)$:

\textbf{(i) Query Complexity ($D_q$):} Intrinsic difficulty score ($0 \le D_q \le 100$) estimated in the scouting phase. It quantifies reasoning depth (e.g., refactoring vs. symbol lookup) and serves as the primary scaler for the compute budget.

\textbf{(ii) Repository Entropy ($H_r$):} Structural complexity factor ($0.5 \le H_r \le 2.0$) derived from topological analysis (e.g., file count, code depth). High entropy implies sparse information density, necessitating a wider exploration horizon.

\textbf{(iii) Resource Consumption ($L_t$):} Cumulative context volume (in lines of code) held at step $t$. This acts as a dynamic cost constraint to strictly enforce the adaptive line budget.

\textbf{(iv) Iteration Depth ($t$):} Current recursive step count ($1 \le t \le T$). It serves as a hard horizon constraint to prevent infinite reasoning loops, overriding budget availability.

\textbf{(v) Epistemic Confidence ($\kappa_t$):} A self-assessed probability metric ($0 \le \kappa_t \le 100$) where the agent evaluates the sufficiency of the current context $L_t$ to answer the query. This functions as the reward signal for the termination policy.

\subsubsection{Adaptive Context Optimization}
\label{sec:optimization}

Guided by the initialized state $S_t$, \model\ employs a policy $\pi(S_t)$ to orchestrate the unit selection trajectory. This policy dynamically balances reasoning accuracy ($\kappa_t$) against token economy ($t$ and $L_t$) via the following mechanisms.

\vspace{-0.1in}
\paragraph{Dynamic Budgeting \& Iteration Control}
The agent first establishes a resource boundary based on the environmental complexity. We assign a dynamic line budget $B \propto D_q \cdot H_r$, ensuring that complex queries in high-entropy repositories are granted expanded computational resources, while simple lookups are constrained to minimize noise.
The iteration flow is governed by the epistemic confidence $\kappa_t$:
\begin{itemize}[leftmargin=*]
    \vspace{-0.1in}
    \item \textbf{Fast Path:} If initial scouting yields high confidence ($\kappa_0 \ge \tau$), the policy bypasses expensive tool execution and query augmentation to minimize latency. Instead, it executes standard retrieval followed by graph extension and immediately proceeds to the refinement phase. This enforces a robust two-round verification process, preventing hallucinations from premature generation.
    \vspace{-0.05in}
    \item \textbf{Iterative Expansion:} If $\kappa_t < \tau$, the agent continues the retrieval loop. The policy monitors the \textit{Information Gain Rate}, defined as $IGR_t = \frac{\kappa_t - \kappa_{t-1}}{L_t - L_{t-1}}$, which quantifies the confidence improvement per unit of context expansion. We accept a single-step confidence drop ($\Delta \kappa_t < 0$) as a corrective adjustment. Termination is triggered only after consecutive low-gain steps ($IGR_t < \epsilon$) or when the projected cost exceeds the budget.
    \vspace{-0.1in}
\end{itemize}
Consequently, iteration stops when one of three conditions is met: (1) \emph{Sufficiency:} $\kappa_t \ge \tau$; (2) \emph{Inefficiency:} Consecutive low IGR indicating negligible returns; or (3) \emph{Exhaustion:} The projected cost exceeds the remaining budget ($B - L_t$).

\vspace{-0.1in}
\paragraph{Priority-Based Context Selection}
To strictly enforce the line budget $B$ while preserving critical information, \model\ selects code units $u$ based on a priority score $P(u)$:
\begin{equation}
    P(u) = w_1 \cdot \text{Rel}(u) + w_2 \cdot \mathbb{I}_{\text{tool}}(u) + w_3 \cdot \text{Density}(u)
\end{equation}
where $w_i$ are weighting factors. $\text{Rel}(u)$ is the relevance score, $\mathbb{I}_{\text{tool}}$ marks elements explicitly found by tools (ensuring correctness), and $\text{Density}(u)$ prioritizes fine-grained units over large files. The system greedily retains the highest-scoring units until the cumulative size reaches $B$.
\section{Evaluation}

\label{sec:eval}
In this section, we evaluate \model\ on a diverse set of repository-scale code reasoning tasks by answering the following six research questions:
\textbf{RQ1:} How does \model\ perform on repo-level QA tasks compared with existing baselines?
\textbf{RQ2:} How effective is \model\ at localizing relevant code in large-scale repositories?
\textbf{RQ3:} How capable is \model\ in performing end-to-end execution of diverse repository tasks?
\textbf{RQ4:} How efficient is \model\ in terms of token usage?
\textbf{RQ5:} Which components contribute most to \model’s performance?
\textbf{RQ6:} How does performance vary across different LLM?

\subsection{Experimental Setup}
\paragraph{Benchmarks} We conduct a comprehensive evaluation of \model\ across three distinct capabilities: repository-level question answering, file localization, and end-to-end tasks. Specifically, we utilize SWE-QA~\cite{peng2025sweqa} and LongCodeQA~\cite{rando2025longcodebench} to assess cross-file reasoning and long-context comprehension; LOC-BENCH (SWE-Bench-Lite subset)~\cite{chen2025locagent} to measure the accuracy of fault localization in real-world maintenance scenarios; and GitTaskBench~\cite{ni2025gittaskbench} to validate execution-based performance on complex, actionable repository tasks. Details are provided in Appendix~\ref{app:exp_setting}.

\vspace{-0.1in}
\paragraph{Baselines \& Settings}
Baselines and settings as follows:

\noindent \textbf{(1) Direct LLM.} We evaluate standard LLMs in two settings:
(i) \textit{Zero-Shot} (for SWE-QA), where the model is prompted without any external repository context; and
(ii) \textit{Full-Context} (for LongCodeQA), where the entire repository content is fed into the model's long-context window.

\noindent \textbf{(2) RAG-based Methods.} We employ standard RAG baselines with varying retrieval granularities:
\textit{Func RAG} retrieves function-level semantic chunks;
\textit{Sliding RAG} retrieves overlapping code spans via a sliding window; and
\textit{File BM25 RAG} retrieves a subset of relevant files based on sparse keyword matching scores.

\begin{table*}[t]
  \centering
  \caption{Comparison of baseline performance on the SWE-QA dataset (3-repository subset).}
  \vspace{-0.05in}
  \setlength{\tabcolsep}{4.5pt}
  \renewcommand{\arraystretch}{1.0}
  \scriptsize
  \resizebox{\textwidth}{!}{%
  \begin{tabular}{@{}c*{11}{c}@{}}
    \toprule
    \multirow{2}{*}{\textbf{Metric}} &
    \multicolumn{5}{c}{\textbf{Community Baselines}} &
    \multicolumn{5}{c}{\textbf{Commercial Tools}} &
    \multicolumn{1}{c}{\textbf{Ours}} \\
    \cmidrule(lr){2-6}\cmidrule(lr){7-11}\cmidrule(lr){12-12}
    & \makecell[c]{Direct LLM}
    & \makecell[c]{Func RAG}
    & \makecell[c]{Sliding RAG}
    & \makecell[c]{SWEQA-Agent}
    & \makecell[c]{}
    & \makecell[c]{DeepWiki}
    & \makecell[c]{CodeWiki}
    & \makecell[c]{Gemini Code}
    & \makecell[c]{Claude Code}
    & \makecell[c]{Cursor}
    & \makecell[c]{\textbf{\model}} \\
    \midrule
    Correctness   & 7.36 & 7.96 & 8.06 & 8.36 &  & 8.02 & 7.89 & 8.03 & 8.31 & 8.30 & 8.44 \\
    Completeness  & 3.71 & 5.75 & 5.99 & 6.94 &  & 5.90 & 6.49 & 6.09 & 7.25 & 8.13 & 7.94 \\
    Clarity       & 7.38 & 8.97 & 9.01 & 9.10 &  & 8.73 & 8.75 & 8.97 & 8.76 & 8.71 & 8.84 \\
    Relevance     & 8.94 & 9.56 & 9.65 & 9.72 &  & 9.36 & 9.21 & 9.68 & 9.56 & 9.39 & 9.35 \\
    Reasoning     & 4.91 & 7.38 & 7.51 & 8.22 &  & 7.52 & 7.82 & 7.92 & 8.21 & 8.64 & 8.72 \\
    \midrule
    \textbf{Total Score}         & 32.30 & 39.62 & 40.22 & 42.33 &  & 39.54 & 40.15 & 40.69 & 42.08 & \underline{43.17} & \textbf{43.28} \\
    \bottomrule
  \end{tabular}%
  \label{tab:sweqa_mainresult}
  \vspace{-0.15in}
  }
\end{table*}

\begin{table}[t]
\centering
\caption{Performance comparison on LongCodeQA benchmark.}
\vspace{-0.05in}
\resizebox{\columnwidth}{!}{%
\begin{tabular}{l|cccccc}
\hline
\textbf{LongCodeQA} & \textbf{32K} & \textbf{64K} & \textbf{128K} & \textbf{256K} & \textbf{512K} & \textbf{1M} \\
\hline
Jamba 1.5 - 400B Large & 69.0 & 69.7 & 72.8 & 54.2 & - & - \\
Llama 3.1 - 405B Ins. & 69.9 & 72.4 & 67.4 & - & - & - \\
Llama 4 Scout & 66.4 & 73.7 & 70.7 & 63.1 & 78.7 & 76.0 \\
GPT-4o & 65.5 & 76.3 & 74.3 & - & - & - \\
GPT-4.1 & 72.6 & 73.7 & 78.3 & 72.3 & 78.7 & 80.0 \\
\hline
Qwen2.5-14B Ins. & 61.9 & \textbf{65.8} & 68.5 & 63.1 & 70.2 & 40.0 \\
+ \textbf{\model} & \textbf{63.7} & 61.8 & \textbf{73.3} & \textbf{63.1} & \textbf{72.3} & \textbf{72.0}  \\
\hline
Gemini 2.5 Pro & 75.2 & 71.1 & 71.7 & 67.7 & 68.1 & 69.8 \\
+ \textbf{\model} & \textbf{77.9} & \textbf{79.0} & \textbf{75.0} & \textbf{75.4} & \textbf{80.9} & \textbf{80.0} \\
\hline
Claude 3.5 Sonnet & 65.5 & 69.7 & 71.7 & 66.6 & - & - \\
+ File BM25 RAG & 25.55 & 31.18 & 24.83 & 25.95 & 17.19 & 12.77 \\
+ \textbf{\model} & \textbf{69.9} & \textbf{69.7} & \textbf{72.83} & \textbf{78.5} & \textbf{76.6} & \textbf{78.0} \\
\hline
\end{tabular}
\label{tab:longcodeqa_mainresult}
\vspace{-0.1in}
}
\end{table}

\textbf{(3) Agent \& Procedure-based Methods.} We evaluate representative agentic and procedural frameworks, including SWEQA-Agent~\cite{peng2025sweqa}, Agentless~\cite{xia2025agentless}, MoatlessTools~\cite{MoatlessTools2025}, SWE-agent~\cite{yang2024sweagent}, OpenHands~\cite{wang2025openhands}, and LocAgent~\cite{chen2025locagent}.

\textbf{(4) Commercial Tools.} We also compare against commercial tools: DeepWiki~\cite{deepwiki2025}, CodeWiki~\cite{hurley2025codewiki}, Gemini Code~\cite{gemini_importcode_2025}, Claude Code~\cite{anthropic2026claudecode}, and Cursor~\cite{cursor2026}.

We adopt standard settings and follow the official metrics established for each benchmark, as detailed in Appendix~\ref{app:exp_setting}.

\subsection{Performance on Repo-QA Tasks (RQ1)}

We evaluate \model's effectiveness on repository-scale QA tasks by comparing it against direct LLM usage, RAG \& agent-based methods, and commercial tools. Specifically, both Community baselines and \model\ use Gemini-3-Flash as the backbone model, while commercial tools are kept in default mode. Table~\ref{tab:sweqa_mainresult} and Table~\ref{tab:longcodeqa_mainresult} present results on SWE-QA and LongCodeQA benchmarks, respectively.

\textbf{Comparison with Direct LLM.} 
\textbf{(i)} \emph{On SWE-QA}, FastCode substantially outperforms Direct LLM (Total 43.28 vs. 32.30), with the largest gains in Completeness (7.94 vs. 3.71) and Reasoning (8.72 vs. 4.91). This shows that repo QA benefits from grounded evidence: FastCode can locate relevant code context instead of relying on parametric knowledge.
\textbf{(ii)} \emph{On LongCodeQA}, FastCode also complements long-context LLM usage: instead of reading everything, FastCode’s targeted acquisition yields consistent improvements across budgets (e.g., Gemini 2.5 Pro improves with FastCode at multiple window sizes). Overall, the results indicate that FastCode’s benefit comes from both finding missing evidence and reducing noise, which is crucial when context windows are finite and distractors are abundant.

\textbf{Comparison with RAG and Agentic Baselines.} 
\model\ substantially outperforms traditional retrieval and agentic approaches, achieving 43.28 on SWE-QA versus Func RAG (39.62), Sliding RAG (40.22), and SWEQA-Agent (42.33). The performance gap is most pronounced in Completeness (7.94 vs. 6.94), where our hierarchical indexing captures semantics at multiple granularities while chunk-based methods fragment logical units, and graph-guided expansion systematically traverses structural relationships that tool-based exploration misses.  
On LongCodeQA, we find that naive file retrieval can be harmful: Claude 3.5 Sonnet with File BM25 RAG drops sharply from 65.5 to 25.55 in 32K subset, likely because lexical, file-level matching both misses transitive dependencies and retrieves large, low-signal files that crowd out key evidence. In contrast, FastCode maintains strong performance, reflecting more precise, dependency-complete context construction. Overall, FastCode consistently improves over RAG and agent baselines by reducing fragmentation and noise under a fixed budget.

\textbf{Comparison with Commercial Tools.}
FastCode surpasses commercial systems spanning wiki-style summarization tools (DeepWiki/CodeWiki), imported-repo Q\&A (Gemini Code), and agentic assistants (Claude Code, Cursor).
We posit that wiki-style tools emphasize broad documentation coverage rather than execution- and symbol-level grounding, while interactive assistants may still over-consume context during exploration. 
Agentic assistants benefit from strong interaction loops, which explains Claude Code and Cursor's competitive performance. Yet our framework demonstrates that competitive performance can also be achieved through alternative means. These findings demonstrate that the proposed framework can precisely locate target context by relying solely on lightweight metadata, thereby matching or exceeding commercial-grade systems.

\subsection{Performance on Localization Tasks (RQ2)}
\begin{figure}[t]
    \centering
    \includegraphics[width=1\columnwidth]{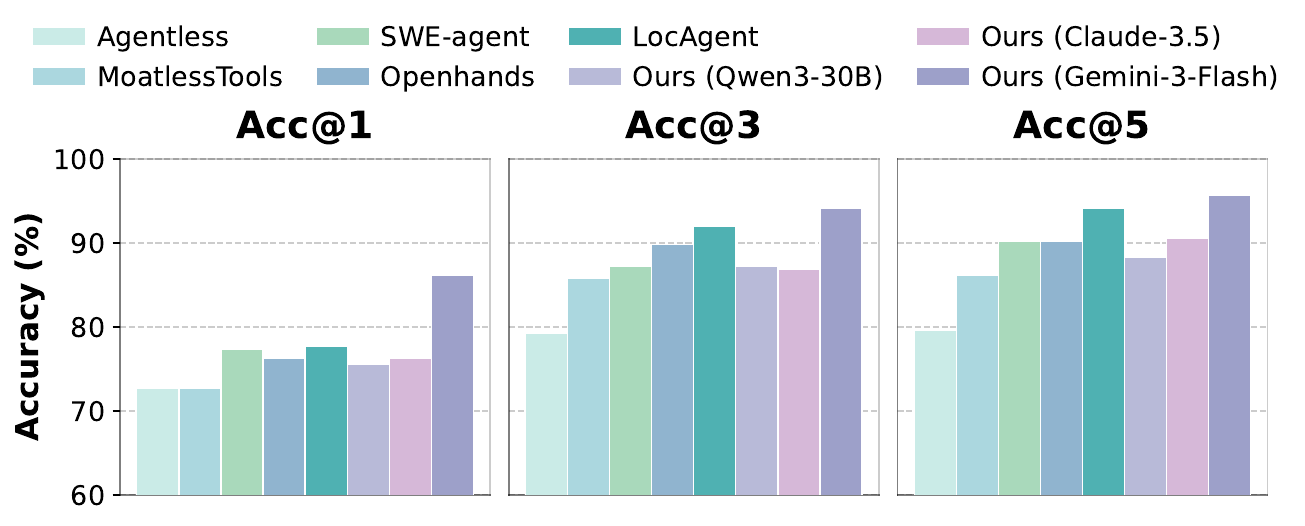}
    \vspace{-0.2in}
    \caption{File-level localization accuracy (Acc@K) on Loc-Bench.}
    \label{fig:locbench_result}
    \vspace{-0.2in}
\end{figure}

Figure~\ref{fig:locbench_result} presents the file-level localization accuracy on SWE-Bench-Lite dataset from LOC-BENCH. Overall, \model\ demonstrates superior capability in identifying relevant files compared to existing agentic frameworks.

\textbf{Superior Performance.} \model\ achieves exceptional performance with Gemini-3-Flash, establishing a new sota across all metrics. Specifically, it attains an Acc@1 of 86.13\%, significantly outperforming the previous best method, LocAgent (77.74\%), by a margin of 8.39\%. This substantial improvement underscores the effectiveness of our approach in accurately pinpointing target files across diverse maintenance scenarios, even in complex repositories.

\textbf{Robustness and Cost-Effectiveness.} Beyond peak performance, our framework demonstrates consistent effectiveness across diverse backends, ranging from the proprietary Claude-3.5 Sonnet to the open-weights Qwen3-Coder-30b-A3b. As shown in Figure~\ref{fig:locbench_result}, \model\ maintains an Acc@1 exceeding 75\% across these different architectures. Notably, the performance with the local Qwen3-Coder-30b-A3b (75.55\%) is highly competitive with—and effectively matches—the results obtained using the much more capable Claude-3.5 Sonnet. This suggests that \model\ is not only model-agnostic but also capable of eliciting high-tier performance from smaller, resource-constrained local models, offering a viable alternative to costly proprietary APIs.

\subsection{Results on End-to-End Repository Tasks (RQ3)}

\begin{figure}[t]
    \centering
    \includegraphics[width=1\columnwidth]{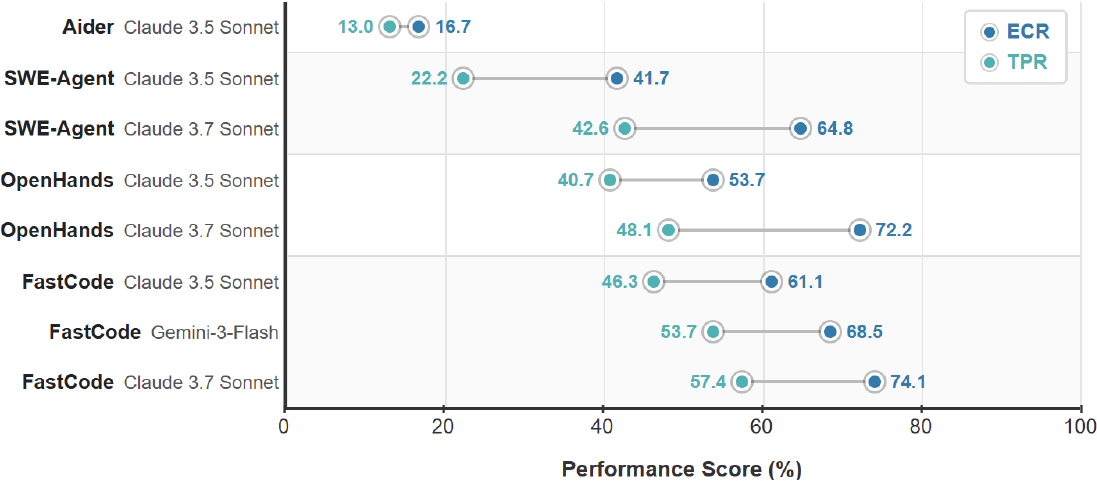}
    \caption{Performance of different frameworks on GitTaskBench.}
    \label{fig:gittask_result}
    \vspace{-0.2in}
\end{figure}

We evaluate \model\ on GitTaskBench to assess its capability in handling complex, real-world repository maintenance scenarios. As shown in the Figure~\ref{fig:gittask_result}, \model\ consistently achieves state-of-the-art performance across varying LLM backbones. Powered by Claude 3.5 Sonnet, our framework attains a Task Pass Rate (TPR) of 46.30\%, surpassing strong baselines like OpenHands (40.74\%) and SWE-Agent (22.23\%). This performance gap widens with the stronger Claude  3.7 Sonnet model, where \model\ reaches a peak TPR of 57.41\% and an Execution Completion Rate (ECR) of 74.07\%. Notably, \model\ demonstrates exceptional resource efficiency: when utilizing the lightweight Gemini-3-Flash backbone, it achieves a TPR of 53.70\%—outperforming the OpenHands framework running on the significantly more powerful Claude 3.7 Sonnet. 
This effectiveness stems from our navigation and context management mechanisms, which decouple exploration from consumption; by pruning the search space via lightweight scouting and dynamically regulating the token budget, \model constructs a precise context that maximizes reasoning accuracy within resource constraints.

\begin{figure*}[htbp]
    \centering
    \begin{subfigure}[b]{0.24\textwidth}
        \centering
        \includegraphics[width=\textwidth]{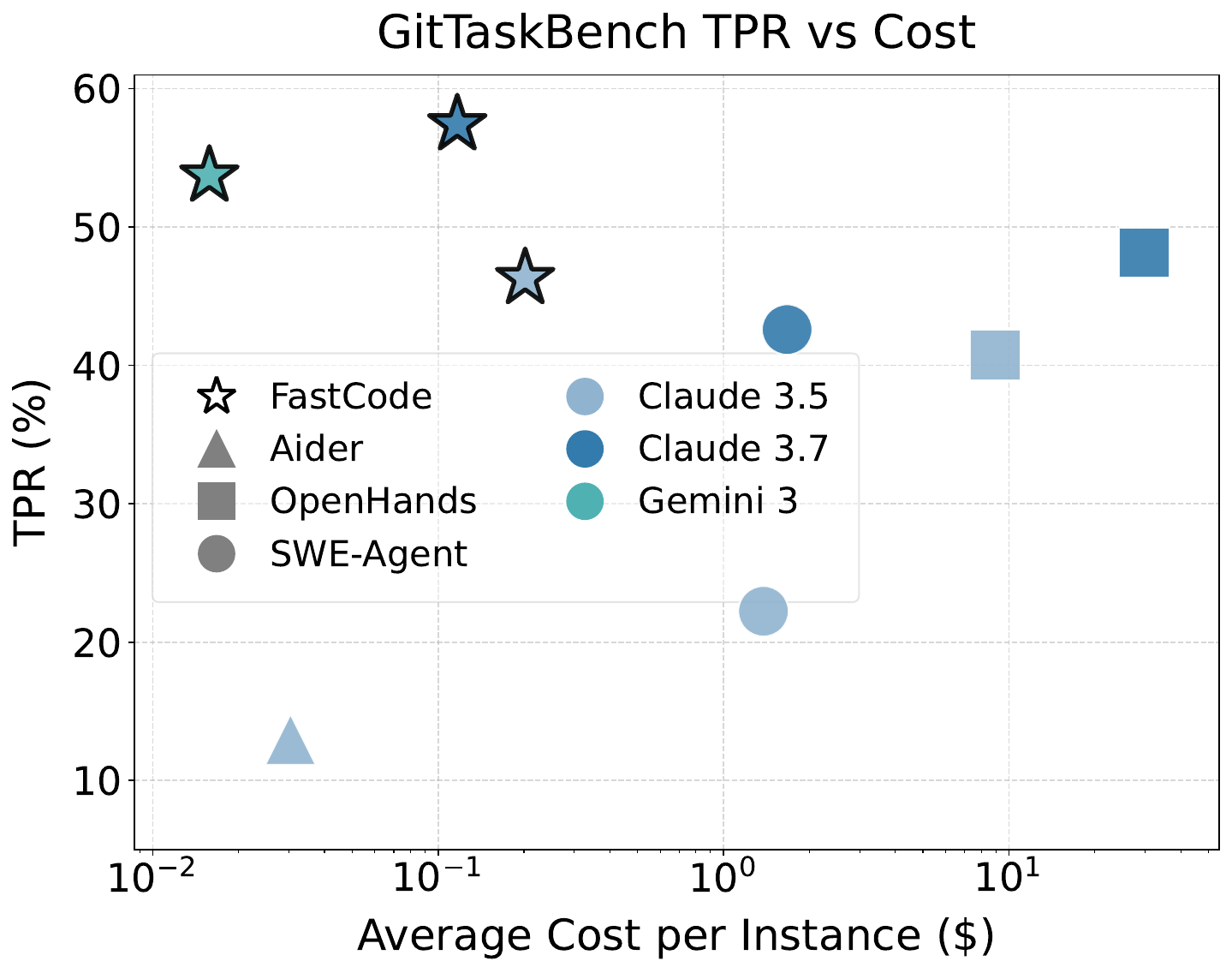}
    \end{subfigure}
    \hfill
    \begin{subfigure}[b]{0.24\textwidth}
        \centering
        \includegraphics[width=\textwidth]{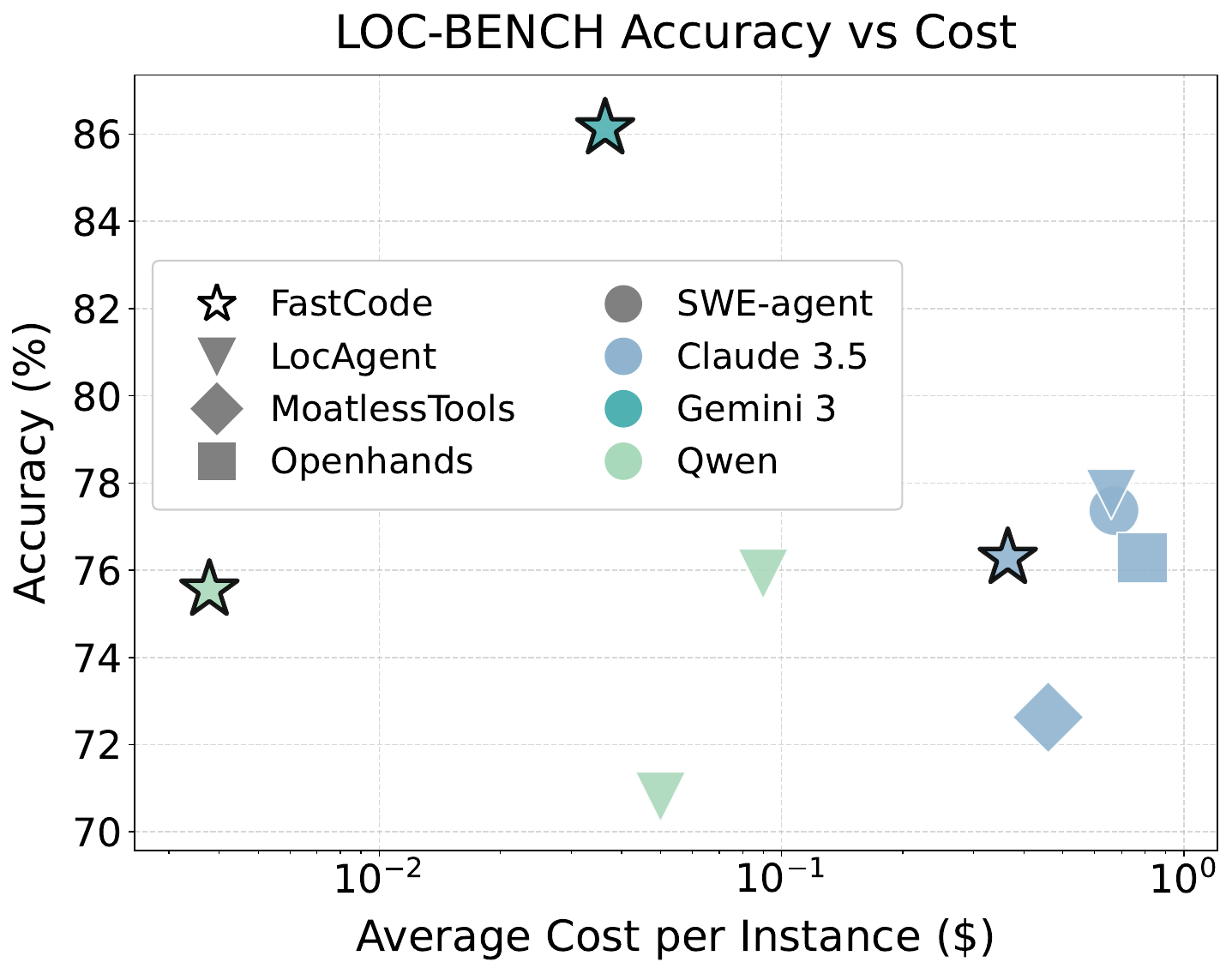}
    \end{subfigure}
    \hfill
    \begin{subfigure}[b]{0.24\textwidth}
        \centering
        \includegraphics[width=\textwidth]{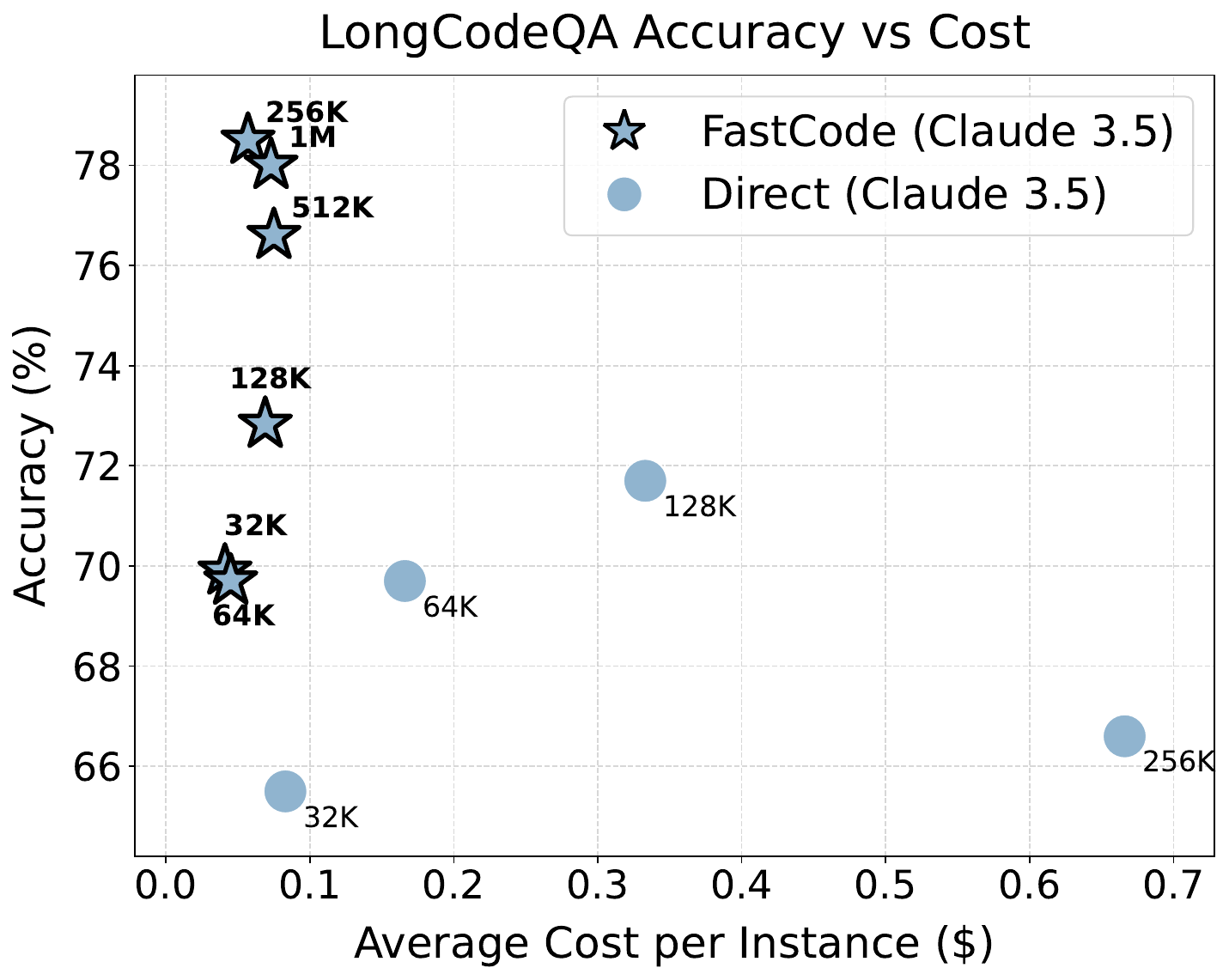}
    \end{subfigure}
    \hfill
    \begin{subfigure}[b]{0.24\textwidth}
        \centering
        \includegraphics[width=\textwidth]{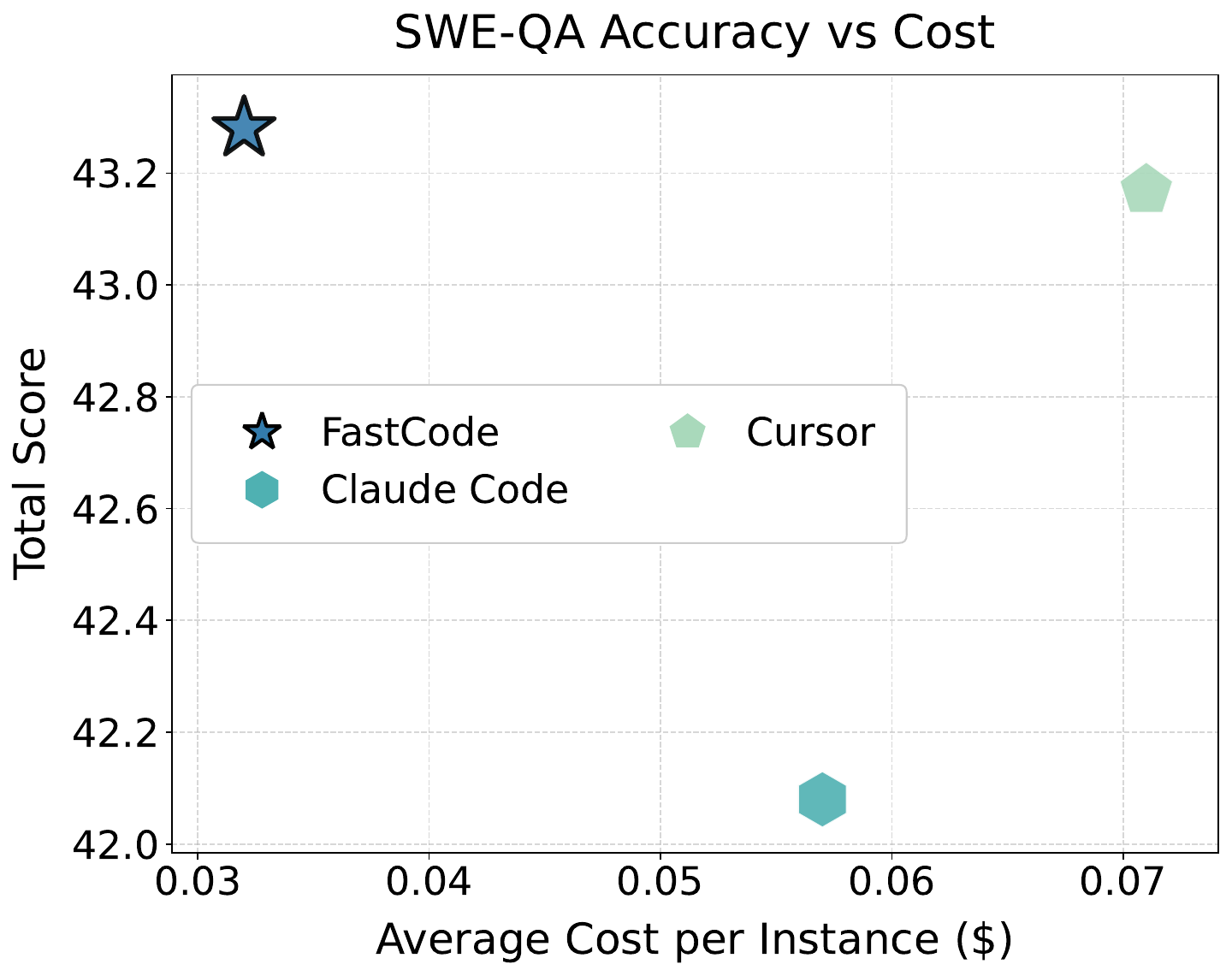}
    \end{subfigure}
    \vspace{-0.05in}
    \caption{Performance-cost tradeoffs for baselines and our framework across four benchmarks.}
    \label{fig:tradeoff}
    \vspace{-0.05in}
\end{figure*}

\subsection{Cost-Performance Trade-off (RQ4)}

As shown in Figure~\ref{fig:tradeoff}, \model\ demonstrates a decisive advantage in cost-efficiency by decoupling information exploration from consumption. Through our structural scouting mechanism, we achieve competitive or superior performance across all tasks while reducing costs by orders of magnitude compared to state-of-the-art baselines.

\textbf{(1) Scalable QA Efficiency.} On SWE-QA, \model\ reduces costs by approximately \textbf{55\%} compared to commercial tools like Cursor while maintaining high accuracy. On LongCodeQA, while baseline costs grow quadratically with context length (escalating to over \$2.60/query), \model\ maintains near-constant costs regardless of input scale. At 256K tokens, this results in a \textbf{$>$90\% cost reduction}, proving that selective context construction effectively neutralizes the "context tax" of long-context models.

\textbf{(2) Localization \& Deployment Viability.} For file localization (LOC-BENCH), \model\ is \textbf{18--22× cheaper} than agentic frameworks like OpenHands and LocAgent. Notably, our open-weights variant (Qwen3-Coder-30b) achieves 97\% of the best baseline's performance at merely \textbf{0.5\% of the cost} (approx. 200× reduction), demonstrating the feasibility of high-accuracy local deployment without reliance on expensive commercial APIs.

\textbf{(3) End-to-End Agentic Efficiency.} In GitTaskBench, the efficiency gap widens further. \model\ consistently outperforms SWE-Agent and OpenHands in TPR while drastically lowering expenses. Depending on the backbone model, our approach ranges from \textbf{15× to over 2000× cheaper} than OpenHands. Specifically, the Gemini-3-Flash variant offers a practical sweet spot, outperforming Aider significantly while reducing costs by three orders of magnitude compared to SOTA agents, making continuous integration and large-scale automated coding economically viable.

\subsection{Ablation Study (RQ5)}
\begin{figure}[t]
    \centering
    \includegraphics[width=\columnwidth]{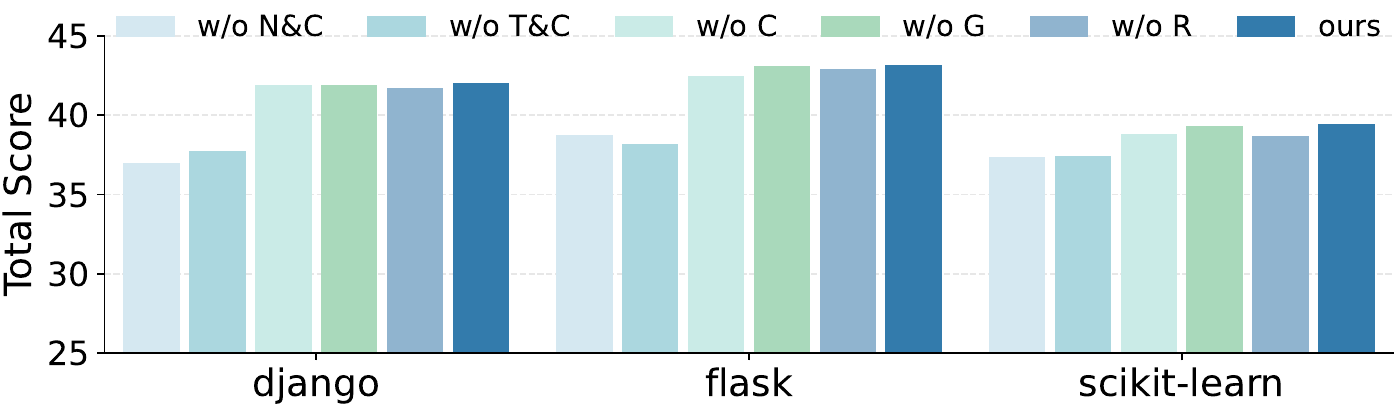}
    \caption{Ablation results on SWE-QA.}
    \label{fig:ablation}
    \vspace{-0.15in}
\end{figure}

To validate the contribution of each component, we conduct ablation experiments on a three-repository subset of SWE-QA using Qwen3-Coder-30b-A3b. As shown in Figure~\ref{fig:ablation}, we evaluate five variants that progressively remove core mechanisms. The results reveal two key findings:

\textbf{Individual components provide complementary value.} The w/o R variant, which removes hybrid retrieval, shows a -0.60\% reduction, highlighting that semantic-lexical matching remains necessary for efficient initial candidate identification even with tool-based exploration. Eliminating the graph extension mechanism (w/o G) causes a -0.83\% drop, indicating that structural dependencies captured by call/inheritance/dependency relations uncover non-lexical relationships missed by retrieval alone. When removing Cost-Aware Context Management (w/o C), the system loses dynamic budget allocation and epistemic confidence monitoring, resulting in a -0.92\% decline as the agent fails to balance exploration depth against resource constraints.

\textbf{Navigation mechanisms are critical.} The w/o N\&C variant, which removes both query augmentation and tool-based exploration (also w/o C), exhibits the most significant performance degradation (-5.06\% average), demonstrating the fundamental importance of bridging the semantic gap between natural language queries and code structure. The w/o T\&C variant, which retains query augmentation but eliminates tool-based scouting, shows a smaller yet substantial drop (-4.31\% average). This gap confirms that while query refinement establishes semantic alignment, structural exploration through lightweight metadata is essential for efficient candidate pruning in large repositories.

\section{Related Work}

\subsection{Retrieval \& Graph-Augmented Code Reasoning}
Repository-scale code reasoning is fundamentally constrained by limited context budgets, motivating RAG as a standard approach for grounding model outputs in relevant code evidence. Early code RAG methods primarily relied on dense semantic retrieval to select code chunks similar to the query~\cite{feng2020codebert, zhang2023repocoder}. However, representing code as flat text can obscure critical structural signals (e.g., imports, call relations, and data/control dependencies), which limits multi-hop reasoning performance~\cite{Ding2023CROSSCODEEVAL}. To better preserve program structure, recent work incorporates graph- or dependency-aware representations for retrieval and context construction~\cite{liu2025codexgraph, liu2024graphcoder, chen2025locagent}. For example, GraphCoder~\cite{liu2024graphcoder} leverages static structure to retrieve structurally related nodes (e.g., $k$-hop neighborhoods) to enrich the context, while LocAgent~\cite{chen2025locagent} combines graph indices with agentic navigation to iteratively localize relevant code.
Despite these advancements, it remains an open challenge to cost-efficiently pinpoint target contexts in complex repositories across diverse query scenarios.

\subsection{Autonomous Coding Agent}
Autonomous coding agents have emerged as a practical approach to repository-scale code reasoning, where a system must localize relevant evidence in a large codebase and execute grounded actions (e.g., debugging, or patching) to satisfy a natural language request~\cite{jimenez2024swebench}. Representative frameworks such as MapCoder~\cite{islam2024mapcoder}, CodeAgent~\cite{zhang2025codeagent}, SWE-agent~\cite{yang2024sweagent}, and OpenHands~\cite{wang2025openhands}, as well as commercial systems like Cursor~\cite{cursor2026} and Claude Code~\cite{anthropic2026claudecode}, typically follow an iterative ``Observe--Think--Act'' paradigm: the model repeatedly invokes environment tools (e.g., bash/Python, search, file open/edit, tests) to interact with the repository, progressively gather context, and perform operations. Despite their effectiveness in repository-level tasks, current agents suffer from high token consumption driven by the context accumulation of multi-turn interactions. In this work, we aim to bridge this efficiency gap by significantly reducing token usage while maintaining competitive reasoning accuracy.
\section{Conclusion}

We introduced \model, a framework that resolves the tension between reasoning accuracy and computational cost by decoupling repository exploration from content consumption. Through a proposed scouting-first paradigm that navigates a semantic-structural map, our approach effectively isolates high-value context without the overhead of iterative full-text ingestion. Extensive evaluations across four benchmarks demonstrate that \model\ consistently outperforms SOTA baselines, achieving superior accuracy with significantly reduced token usage. These results validate that \model\ offers a scalable and resource-efficient solution for complex, repository-scale software engineering tasks.

\clearpage



\balance
\bibliography{example_paper}
\bibliographystyle{icml2026}


\clearpage
\nobalance
\appendix

\section{Appendix}

This appendix provides supplementary implementation details and extended experimental analysis to support the findings presented in the main text. The content is organized as follows. Section~\ref{app:sec_a} details the algorithms and construction process for Symbol-Aware Relation Modeling. Section~\ref{app:exp_setting} outlines the Experimental Setup, including specific hyperparameters, baseline configurations, and metric definitions. Finally, Section~\ref{app:add_exp} presents Additional Experiments and analysis that further validate the robustness and effectiveness of our framework.

\subsection{Details of Symbol-Aware Relation Modeling}
\label{app:sec_a}
To instantiate $\mathcal{G}$ with high precision, we employ a two-stage resolution mechanism that operates on the static syntax trees. First, for structural connectivity ($G_{dep}$ and $G_{inh}$), we implement Global Symbol Resolution, which utilizes a deterministic module map to resolve dynamic import patterns---including aliased namespaces (e.g., \texttt{import numpy as np}) and relative paths---to unique definition IDs. Second, for execution logic ($G_{call}$), we mitigate Python's dynamic typing ambiguity via Scope-Constrained Type Inference. By tracking variable assignments within specific AST scopes (e.g., constructor initialization), our system infers instance types to link method calls (e.g., \texttt{self.loader.load()}) directly to their target definitions.

\subsection{Experimental Setup}
\label{app:exp_setting}
\paragraph{Benchmark}
We evaluate our \model\ framework on the following three classes of repository-centric benchmarks:

\textbf{(1) Repository-level Code QA.}
SWE-QA~\cite{peng2025sweqa} is a repository-level question answering benchmark with 576 human-validated QA pairs from 12 widely used open-source Python repositories, targeting holistic codebase understanding that requires cross-file, multi-hop reasoning (e.g., intent inference, dependency tracing, and feature localization). LongCodeQA~\cite{rando2025longcodebench}, a subset of LongCodeBench, complements this setting with multiple-choice questions derived from real GitHub issue discussions, emphasizing long-context repository comprehension under realistic problem statements. It contains 443 instances spanning input token lengths from 32K to 1M tokens.

\textbf{(2) File Localization.}
We further evaluate on SWE-Bench-Lite from LOC-BENCH~\cite{chen2025locagent}, which measures a model’s ability to localize the relevant files and code regions for resolving real-world issues. Built from 274 issue--patch pairs in popular Python projects, this benchmark stresses accurate fault/feature localization as a prerequisite to downstream patch generation, covering diverse maintenance scenarios such as bug fixes and feature updates.

\textbf{(3) End-to-End Repository Tasks.}
GitTaskBench~\cite{ni2025gittaskbench} evaluates agentic performance on end-to-end tasks that require setting up environments, navigating existing GitHub codebases, and producing runnable outputs under practical success criteria. It contains 54 human-curated tasks across multiple modalities/domains, each paired with an automated evaluation harness (e.g., execution-based completion and pass rate) and cost-aware metrics to quantify effectiveness under realistic resource constraints.

\paragraph{Baselines}
Details of Agent \& Procedure-based Methods and Commercial Tools are as follows:



\noindent \textbf{(i) Agent \& Procedure-based Methods.} We include agentic and procedural baselines for repository-level reasoning. SWEQA-Agent~\cite{peng2025sweqa} is a ReAct-style tool-using agent that iteratively searches and reads the repository. Agentless~\cite{xia2025agentless} is a lightweight, procedure-driven LLM pipeline for repository-level reasoning that follows a structured workflow. MoatlessTools
~\cite{MoatlessTools2025}, SWE-agent~\cite{yang2024sweagent}, and OpenHands~\cite{wang2025openhands} represent tool-augmented agents that gather code context through repository actions such as search, file navigation, and command execution; LocAgent~\cite{chen2025locagent} performs graph-guided code localization by constructing a heterogeneous repository graph.

\noindent \textbf{(ii) Commercial Tools.} We evaluate commercial repository assistants as baselines for documentation and interactive code understanding. DeepWiki~\cite{deepwiki2025} and CodeWiki~\cite{hurley2025codewiki} ingest a repository and generate wiki-style documentation (e.g., project overview, modules, and APIs). Gemini Code (import code)~\cite{gemini_importcode_2025} supports repository import and provides interactive natural-language Q\&A grounded in the imported codebase. Claude Code~\cite{anthropic2026claudecode} is a proprietary coding assistant for repository-level workflows such as code understanding and refactoring. Cursor~\cite{cursor2026} is a commercial IDE with integrated AI assistance that supports project-wide context and in-editor editing workflows.

\paragraph{Configurations} For SWE-QA, we evaluate community baselines and \model\ with Gemini-3-Flash. For DeepWiki and CodeWiki, we obtain answers via their default interaction workflow. Additionally, we use GPT-5.2 as the evaluation model to assess the performance of all baselines and our method on SWE-QA. For Gemini Code and Claude Code, we use Gemini-3-Pro and Claude Sonnet 4.5 as the backbone models, respectively, while Cursor is run in Auto mode to select the model automatically. For LongCodeQA and LOC-BENCH, we extract the repository associated with each question and provide it as input to FastCode for reasoning. For GitTaskBench, we prompt FastCode to generate an executable .sh script and run it using a predefined automation script, and the benchmark scores the run against a unified success criterion.

\paragraph{Cost Calculation} We primarily employ Poe\footnote{https://poe.com/api} and OpenRouter\footnote{https://openrouter.ai/} as API endpoints. Specifically, we access Qwen3-Coder-30B via OpenRouter, while utilizing Poe for all other models. Costs are calculated based on the official pricing (\textit{i.e.}, cost per 1M tokens). Detailed pricing is listed in Table~\ref{tab:api_price}. For baselines, we directly adopt the cost figures reported in their original publications~\cite{chen2025locagent} or relevant benchmark papers~\cite{rando2025longcodebench, ni2025gittaskbench}.

\begin{table}[t]
    \centering
    \caption{Detailed API usage costs. Prices are denoted in USD per 1M tokens.}
    \label{tab:api_price}
    \begin{tabular}{cccc}
        \toprule
        \multirow{2}{*}{\textbf{Model}} & \multirow{2}{*}{\textbf{Provider}} & \multicolumn{2}{c}{\textbf{Price (\$ / 1M)}} \\
        \cmidrule(lr){3-4}
         & & \textbf{Input} & \textbf{Output} \\
        \midrule
        Gemini-3-Flash & Poe & 0.40 & 2.40 \\
        Gemini-2.5-Pro & Poe & 0.87 & 7.00 \\
        Claude 3.5 Sonnet & Poe & 2.60 & 13.00 \\
        Claude 3.7 Sonnet & Poe & 2.60 & 13.00 \\
        Qwen3-Coder-30B & OpenRouter & 0.07 & 0.27 \\
        \bottomrule
    \end{tabular}
    \vspace{-0.15in}
\end{table}

\paragraph{Metrics} For the four benchmarks, we follow the official evaluation metrics established for each task: (i) \emph{SWE-QA}: We evaluate models with an LLM-as-judge rubric that scores each prediction against the reference answer on five dimensions—correctness, completeness, relevance, clarity, and reasoning quality—with a maximum score of 10 for each. (ii) \emph{LongCodeQA}: Performance is evaluated via multiple-choice accuracy (1-of-4 options) across varying context-length intervals. (iii) \emph{LOC-BENCH (SWE-Bench-Lite)}: We report Acc@$k$ ($k \in \{1, 3, 5\}$), where correctness requires all ground-truth files to be present in the top-$k$ candidates. (iv) \emph{GitTaskBench}: We evaluate performance using Execution Completion Rate (ECR), checking for the generation of valid and non-empty outputs, and Task Pass Rate (TPR), which verifies if the outputs satisfy specific quality thresholds and functional requirements tailored to each task.

\begin{table}[h]
\centering
\caption{SWE-QA Performance Comparison}
\label{apptab:swe-qa}
\resizebox{\columnwidth}{!}{%
\begin{tabular}{lcc}
\toprule
\textbf{Method} & \textbf{Total Score} & \textbf{Cost} \\
\midrule
Claude Code & 42.08 & 0.057 \\
Cursor & 43.17 & 0.071 \\
\textbf{FastCode} & \textbf{43.28} & \textbf{0.032} \\
\midrule
\textit{Improvement over Claude Code} & +2.85\% & \textcolor{blue}{-43.86\%} \\
\bottomrule
\end{tabular}
\vspace{-0.15in}
}
\end{table}

\begin{table}[h]
\centering
\caption{LongCodeQA Performance: Direct vs. FastCode-Enhanced}
\label{apptab:longcodeqa}
\resizebox{\columnwidth}{!}{%
\begin{tabular}{lccccc}
\toprule
\multirow{2}{*}{\textbf{Context Length}} & \multicolumn{2}{c}{\textbf{Accuracy (\%)}} & \multicolumn{2}{c}{\textbf{Cost}} & \multirow{2}{*}{\textbf{Acc. Gain}} \\
\cmidrule(lr){2-3} \cmidrule(lr){4-5}
& Direct & +FastCode & Direct & +FastCode & \\
\midrule
32K & 65.5 & \textbf{69.9} & 0.083 & 0.041 & \textcolor{blue}{+6.72\%} \\
64K & 69.7 & 69.7 & 0.166 & 0.045 & +0.00\% \\
128K & 71.7 & \textbf{72.83} & 0.333 & 0.069 & \textcolor{blue}{+1.58\%} \\
256K & 66.6 & \textbf{78.5} & 0.666 & 0.057 & \textcolor{blue}{+17.87\%} \\
512K & - & \textbf{76.6} & 1.331 & 0.075 & - \\
1M & - & \textbf{78.0} & 2.600 & 0.073 & - \\
\midrule
\multicolumn{6}{l}{\textit{Average Cost Reduction: \textbf{-83.01\%} (for comparable lengths)}} \\
\bottomrule
\end{tabular}
\vspace{-0.15in}
}
\end{table}

\begin{table}[h]
\centering
\caption{LOC-BENCH (SWE-Bench-Lite) Performance}
\label{apptab:loc-bench}
\resizebox{\columnwidth}{!}{%
\begin{tabular}{lcccc}
\toprule
\textbf{Method} & \textbf{Acc@1 (\%)} & \textbf{Acc@3 (\%)} & \textbf{Acc@5 (\%)} & \textbf{Cost} \\
\midrule
\multicolumn{5}{l}{\textit{Existing Tools (Claude 3.5 Sonnet)}} \\
\quad MoatlessTools & 72.63 & 85.77 & 86.13 & 0.46 \\
\quad SWE-agent & 77.37 & 87.23 & 90.15 & 0.67 \\
\quad Openhands & 76.28 & 89.78 & 90.15 & 0.79 \\
\quad LocAgent & 77.74 & 91.97 & 94.16 & 0.66 \\
\midrule
\multicolumn{5}{l}{\textit{LocAgent (Fine-tuned Qwen)}} \\
\quad Qwen2.5-7B (ft) & 70.80 & 84.67 & 88.32 & 0.05 \\
\quad Qwen2.5-32B (ft) & 75.91 & 90.51 & 92.70 & 0.09 \\
\midrule
\multicolumn{5}{l}{\textit{FastCode Variants}} \\
\quad Qwen3-coder-30B & 75.55 & 87.23 & 88.32 & \textbf{0.0038} \\
\quad Claude 3.5 Sonnet & 76.28 & 86.86 & 90.51 & 0.3650 \\
\quad \textbf{Gemini-3-Flash} & \textbf{86.13} & \textbf{94.16} & \textbf{95.62} & 0.0364 \\
\midrule
\multicolumn{5}{l}{\textit{vs. Best Baseline (LocAgent + Claude 3.5): Performance Retention / Cost Reduction}} \\
\quad Qwen3-coder-30B & 97.18\% & 94.85\% & 93.80\% & \textcolor{blue}{\textbf{-99.42\%}} \\
\quad Claude 3.5 Sonnet & 98.12\% & 94.44\% & 96.12\% & \textcolor{blue}{-44.70\%} \\
\quad \textbf{Gemini-3-Flash} & \textcolor{blue}{\textbf{110.80\%}} & \textcolor{blue}{\textbf{102.38\%}} & \textcolor{blue}{\textbf{101.55\%}} & \textcolor{blue}{-94.48\%} \\
\bottomrule
\end{tabular}
}
\end{table}

\begin{table}[ht]
\centering
\caption{GitTaskBench Performance Comparison}
\label{apptab:gittaskbench}
\resizebox{\columnwidth}{!}{%
\begin{tabular}{lcccc}
\toprule
\textbf{Method} & \textbf{ECR (\%)} & \textbf{TPR (\%)} & \textbf{Cost} & \textbf{Cost Ratio} \\
\midrule
Aider (Claude 3.5) & 16.67 & 12.96 & 0.0304 & 1.00$\times$ \\
\midrule
SWE-Agent (Claude 3.5) & 41.67 & 22.23 & 1.38 & 45.39$\times$ \\
SWE-Agent (Claude 3.7) & 64.81 & 42.59 & 1.67 & 54.93$\times$ \\
\midrule
OpenHands (Claude 3.5) & 53.70 & 40.74 & 8.95 & 294.41$\times$ \\
OpenHands (Claude 3.7) & 72.22 & 48.15 & 29.8 & 980.26$\times$ \\
\midrule
FastCode (Claude 3.5) & 61.11 & 46.30 & 0.0878 & 2.89$\times$ \\
FastCode (Claude 3.7) & \textbf{74.07} & \textbf{57.41} & 0.1015 & 3.34$\times$ \\
FastCode (Gemini-3-Flash) & 68.52 & 53.70 & \textbf{0.0126} & \textbf{0.41}$\times$ \\
\bottomrule
\end{tabular}
    \vspace{-0.15in}
}
\vspace{0.2mm}

\resizebox{\columnwidth}{!}{%
\begin{tabular}{lcc}
\toprule
\textbf{Comparison} & \textbf{ECR Improvement} & \textbf{Cost Reduction} \\
\midrule
FastCode (3.7) vs. OpenHands (3.7) & +2.56\% & \textcolor{blue}{-99.66\%} \\
FastCode (3.5) vs. OpenHands (3.5) & +13.80\% & \textcolor{blue}{-99.02\%} \\
FastCode (Gemini) vs. OpenHands (3.5) & +27.59\% & \textcolor{blue}{-99.86\%} \\
\bottomrule
\end{tabular}
 \vspace{-0.15in}
}
\end{table}


\subsection{Additional Experiment}
\label{app:add_exp}
\subsubsection{Detail of Cost-Performance Trade-off}

A critical advantage of \model\ lies in its ability to achieve competitive or superior accuracy while substantially reducing costs. This efficiency stems from our structural scouting mechanism and adaptive context management, which decouple exploration from consumption and optimize resource allocation based on dynamic reasoning states. The results are summarized in Table~\ref{apptab:swe-qa}, ~\ref{apptab:longcodeqa}, ~\ref{apptab:loc-bench}, ~\ref{apptab:gittaskbench}.

\textbf{(1) Repo QA Efficiency. }
On SWE-QA, \model\ achieves a total score of 43.28 with an average cost of \$0.032, outperforming both Claude Code (42.08, \$0.057) and Cursor (43.17, \$0.071). This represents a \textbf{55\% cost reduction} compared to Cursor while maintaining competitive accuracy, validating that metadata-driven scouting effectively prunes irrelevant code before expensive content reading.

The efficiency advantage becomes more pronounced as repository size increases. On LongCodeQA, direct input approaches suffer from quadratic cost growth with context length—Claude 3.5 Sonnet's cost escalates from \$0.083 (32K) to \$2.600 (1M tokens). In contrast, \model\ maintains near-constant costs (\$0.041-\$0.073) across all scales while achieving superior accuracy. At the 256K context length, our approach delivers \textbf{91.4\% cost reduction} with a \textbf{17.9\% accuracy improvement}, demonstrating that selective context construction outperforms exhaustive reading.

\textbf{(2) File Localization Efficiency. }On LOC-BENCH, \model\ with Gemini-3-Flash achieves state-of-the-art Acc@1 at only \$0.036 per instance—\textbf{21.9× cheaper} than OpenHands and \textbf{18.3× cheaper} than LocAgent, both using Claude 3.5 Sonnet. Even when paired with Claude 3.5 Sonnet, our framework reduces costs by 45\% compared to LocAgent while maintaining competitive accuracy. Notably, our Qwen3-Coder-30b-A3b variant retains 97\% of the best baseline's accuracy (LocAgent with Claude 3.5 Sonnet) at merely \$0.0038 per instance, enabling nearly \textbf{200× cost reduction} compared to commercial API-based systems—a critical advantage for large-scale deployment.

\textbf{End-to-End Task Efficiency}
GitTaskBench results further validate our efficiency gains in realistic agentic workflows. \model\ with Claude 3.5 achieves 61.11\% ECR and 46.30\% TPR at \$0.088 per task—\textbf{15.7× cheaper} than SWE-Agent (\$1.38) and \textbf{102× cheaper} than OpenHands (\$8.95), while delivering superior task completion rates (46.30\% vs. 22.23\% and 40.74\% TPR). With Claude 3.7, our system achieves 74.07\% ECR and 57.41\% TPR at \$0.102—\textbf{16.5× cheaper} than SWE-Agent (\$1.67) and \textbf{294× cheaper} than OpenHands (\$29.8), while maintaining competitive or superior performance compared to baselines. The Gemini-3-Flash variant (68.52\% ECR, 53.70\% TPR, \$0.013) represents the most cost-effective solution, enabling practical deployment at \textbf{710× lower cost} than OpenHands with Claude 3.5 and \textbf{2,365× lower cost} than OpenHands with Claude 3.7, while significantly outperforming Aider (16.67\% ECR, 12.96\% TPR).

\subsubsection{Evaluation Across LLM Backbones (RQ6)}
To assess the universality of our framework, we evaluate \model\ across a spectrum of LLMs ranging from lightweight edge models (e.g., 3B parameters) to frontier cloud models. As detailed in Table~\ref{tab:different_backbones}, \model\ consistently yields significant performance improvements over both the \textit{Direct LLM} and \textit{SWEQA-Agent} baselines, demonstrating its robustness and model-agnostic nature.

\textbf{Empowering Small Language Models.} A key finding is our framework's ability to bridge the capability gap for smaller models. By providing high-precision, structure-aware context, \model\ enables the \textit{Ministral-3b} model to achieve a score of 36.13, effectively outperforming the much larger \textit{Qwen3-Coder-30b-A3b} in the Direct setting (35.77). This suggests that superior context construction can compensate for lower parameter counts, proving that \model\ is highly effective for resource-constrained, on-device repository reasoning where massive models are deployed.

\textbf{Superiority Over Standard Agents.} While the \textit{SWEQA-Agent} baseline improves performance via iterative tool use, it lacks the global navigation map and adaptive scouting mechanism central to our approach, which allow the model to systematically locate relevant contexts. This is particularly evident with \textit{Qwen3-Coder-30b-A3b}: while the standard agent yields only a marginal gain (+1.31) over the direct baseline, \model\ delivers a substantial boost (+5.67). This indicates that even for models already fine-tuned for code, the bottleneck remains context quality rather than reasoning capability. \model\ unlocks the full potential of these strong backbones by replacing noisy, unstructured retrieval with precise, graph-guided navigation, ensuring consistent state-of-the-art performance across all model sizes.

\begin{table}[t]
\caption{\model\ performance on SWE-QA under different LLMs.}
\centering
\resizebox{\columnwidth}{!}{%
\begin{tabular}{cccc}
\toprule
\textbf{Base Model} & \textbf{Direct LLM} & \textbf{SWEQA-Agent} & \textbf{\model} \\
\midrule
Ministral-3b       & 27.39 & 33.45 & \textbf{36.13} \\
Deepseek-r1-8b     & 30.11 & 34.46 & \textbf{38.56} \\
Qwen3-Coder-30b-A3b    & 35.77 & 37.08 & \textbf{41.44} \\
Gemini-3-Flash     & 32.65 & 42.47 & \textbf{43.44} \\
\bottomrule
\end{tabular}%
\label{tab:different_backbones}
\vspace{-0.15in}
}
\end{table}
\begin{table}[ht]
\caption{Direct LLM results on SWE-QA across all repositories with different backbone LLMs.}
\centering
\resizebox{\columnwidth}{!}{%
\begin{tabular}{lcccc}
\toprule
\textbf{Repository} & \textbf{DeepSeek-R1-8B} & \textbf{Gemini-3-Flash} & \textbf{Ministral-3B} & \textbf{Qwen3-Coder-30B} \\
\midrule
astropy        & 27.58 & 32.42 & 25.60 & 34.08 \\
sphinx         & 28.54 & 32.67 & 24.60 & 34.44 \\
matplotlib     & 31.52 & 33.88 & 28.38 & 37.48 \\
pylint         & 28.46 & 32.79 & 27.33 & 35.56 \\
flask          & 33.46 & 32.12 & 27.81 & 36.42 \\
pytest         & 28.54 & 32.50 & 26.75 & 35.46 \\
django         & 33.31 & 33.31 & 30.90 & 37.56 \\
xarray         & 29.46 & 32.71 & 27.98 & 35.35 \\
scikit-learn   & 31.48 & 31.46 & 27.25 & 35.10 \\
requests       & 31.85 & 34.06 & 31.48 & 38.58 \\
sympy          & 30.33 & 32.29 & 24.15 & 34.73 \\
sqlfluff       & 26.75 & 31.62 & 26.44 & 34.44 \\
\midrule
\textbf{Overall Average} & \textbf{30.11} & \textbf{32.65} & \textbf{27.39} & \textbf{35.77} \\
\bottomrule
\end{tabular}%
\label{tab:direct-llm-repo-results}
\vspace{-0.15in}
}
\end{table}
\begin{table}[ht]
\caption{SWEQA-Agent results on SWE-QA across all repositories with different backbone LLMs.}
\centering
\resizebox{\columnwidth}{!}{%
\begin{tabular}{lcccc}
\toprule
\textbf{Repository} & \textbf{DeepSeek-R1-8B} & \textbf{Gemini-3-Flash} & \textbf{Ministral-3B} & \textbf{Qwen3-Coder-30B} \\
\midrule
astropy        & 31.44 & 41.32 & 35.15 & 36.21 \\
sphinx         & 32.71 & 42.08 & 29.35 & 38.40 \\
matplotlib     & 35.12 & 43.48 & 33.75 & 37.67 \\
pylint         & 34.29 & 42.60 & 34.85 & 37.12 \\
flask          & 35.71 & 42.15 & 38.67 & 37.60 \\
pytest         & 33.25 & 42.92 & 27.44 & 39.77 \\
django         & 35.75 & 43.31 & 31.00 & 29.77 \\
xarray         & 35.23 & 41.17 & 36.21 & 38.62 \\
scikit-learn   & 35.06 & 41.54 & 35.46 & 37.21 \\
requests       & 39.08 & 43.50 & 31.58 & 38.90 \\
sympy          & 34.06 & 43.29 & 35.40 & 37.08 \\
sqlfluff       & 31.79 & 42.27 & 32.56 & 36.60 \\
\midrule
\textbf{Overall Average} & \textbf{34.46} & \textbf{42.47} & \textbf{33.45} & \textbf{37.08} \\
\bottomrule
\end{tabular}%
\label{tab:sweqa-agent-repo-results}
\vspace{-0.15in}
}
\end{table}
\begin{table}[ht]
\caption{FastCode results on SWE-QA across all repositories with different backbone LLMs.}
\centering
\resizebox{\columnwidth}{!}{%
\begin{tabular}{lcccc}
\toprule
\textbf{Repository} & \textbf{DeepSeek-R1-8B} & \textbf{Gemini-3-Flash} & \textbf{Ministral-3-3B} & \textbf{Qwen3-Coder-30B} \\
\midrule
astropy        & 36.40 & 41.58 & 33.88 & 40.33 \\
sphinx         & 37.81 & 44.10 & 35.96 & 41.75 \\
matplotlib     & 40.56 & 43.44 & 38.73 & 41.79 \\
pylint         & 37.98 & 43.42 & 34.06 & 40.65 \\
flask          & 40.27 & 44.98 & 38.12 & 43.19 \\
pytest         & 38.65 & 44.12 & 35.00 & 41.96 \\
django         & 38.50 & 44.06 & 35.25 & 42.04 \\
xarray         & 37.35 & 43.04 & 35.56 & 40.65 \\
scikit-learn   & 36.73 & 41.81 & 34.33 & 39.44 \\
requests       & 42.19 & 44.29 & 41.08 & 43.56 \\
sympy          & 37.04 & 42.77 & 34.62 & 39.79 \\
sqlfluff       & 39.23 & 43.69 & 36.94 & 42.08 \\
\midrule
\textbf{Overall Average} & \textbf{38.56} & \textbf{43.44} & \textbf{36.13} & \textbf{41.44} \\
\bottomrule
\end{tabular}%
\label{tab:fastcode-repo-results}
\vspace{-0.15in}
}
\end{table}

\subsubsection{Full Results on SWE-QA}
To complement the backbone-level summary in Table~\ref{tab:different_backbones}, Tables~\ref{tab:direct-llm-repo-results}, \ref{tab:sweqa-agent-repo-results}, and \ref{tab:fastcode-repo-results} further break down the total score over each of the 12 SWE-QA repositories under \textit{Direct LLM}, \textit{SWEQA-Agent}, and our method, \model. The gains from FastCode are not driven by a small subset of projects: for every backbone, it improves over the direct baseline on all repositories, spanning scientific libraries (e.g., \textit{astropy}, \textit{sympy}), developer tooling (e.g., \textit{pylint}, \textit{sqlfluff}), and web frameworks (e.g., \textit{flask}, \textit{django}). \textit{SWEQA-Agent} is often helpful, but its per-repository behavior is less stable; for example, with \textit{Qwen3-Coder-30B} on \textit{django} it underperforms the direct baseline (29.77 vs.\ 37.56), while FastCode reaches 42.04. Overall, the repository-level results reinforce the conclusion above: improving context construction yields reliable benefits across diverse codebases and model scales.




\end{document}